\documentclass[aps,prb,amsmath,amssymb,reprint,superscriptaddress,preprintnumbers,showpacs,intlimits]{revtex4-1}
\usepackage{bm,mathrsfs}
\usepackage[mathcal]{euscript}
\usepackage{hyperref}
\usepackage{graphicx}
\usepackage{subfig}
\usepackage{color}
\renewcommand{\vec}[1]{\bm{#1}}
\begin{document}

\title{Bloch Point Structure in a Magnetic Nanosphere}

\author{Oleksandr V. Pylypovskyi}
\affiliation{Taras Shevchenko National University of Kiev, 01601 Kiev, Ukraine}

\author{Denis D. Sheka}
\email{sheka@univ.net.ua}
\affiliation{Taras Shevchenko National University of Kiev, 01601 Kiev, Ukraine}
\affiliation{Institute for Theoretical Physics, 03143 Kiev, Ukraine}

\author{Yuri Gaididei}
\affiliation{Institute for Theoretical Physics, 03143 Kiev, Ukraine}

\date{\today}

%
%


\begin{abstract}
A Bloch Point singularity can form a metastable state in a magnetic nanosphere. We classify possible types of Bloch points and derive analytically the shape of magnetization distribution of different Bloch points. We show that external gradient field can stabilize the Bloch point: the shape of the Bloch point becomes radial--dependent one. We compute the magnetization structure of the nanosphere, which is  in a good agrement with performed spin--lattice simulations.
\end{abstract}




\maketitle

\section*{Introduction}

Topological singularities are widely recognized as key to understanding the behavior of wide variety condensed matter systems. Linear topological singularities such as dislocations, disclinations, and vortices, play a crucial role in low--dimensional phase transitions, \cite{Domb83,*Strandburg91,*Yonezawa83} crystalline ordering on curved surfaces \cite{Bowick09}, rotating trapped Bose--Einstein condesates \cite{Fetter09} etc. Recent advances in micro--structuring technology have made it possible to fabricate various nanoparticles with well--prescribed geometry. Much recent research in this field has focused on the statics and dynamics of topological singularities in nanoscale confined systems: essentially inhomogeneous states can be realized in magnetic nanoparticles \cite{Hubert98,Stohr06,Guimaraes09} and ferroelectric nanoparticles \cite{Naumov04}. As a result of the competition between exchange and magnetic dipole--dipole interactions the ground state of magnetic disks with sizes larger than some tens of nanometers is a flux--closure vortex state.

Besides linear singularities there exist also so--called point singularities such as monopoles, Bloch points, boojums. For example, hedgehog (monopole) singularities play a crucial role in the behavior of matter near quantum phase transitions that are seen in a variety of experimentally relevant two--dimensional antiferromagnets, \cite{Senthil04} boojums are relevant in superfluid He--3, \cite{Mermin81,*Volovik03} Bloch points along with Bloch lines are principle in understanding of magnetic bubble dynamics.\cite{Malozemoff79,Hubert98}

The concept of point singularities was introduced in magnetism by \citet{Feldtkeller65b}, who considered different magnetization distributions inside the singularity and proposed first estimations of the Bloch point shape. Later \citet{Doering68} studies how magnetostatic energy governs the Bloch point structure by selecting the rotation angle inside the Bloch point. Bloch point singularities were directly observed in yttrium iron garnet crystals. \cite{Kabanov89} During the last decade Bloch points were also studied by micromagnetic simulations in nanowires, \cite{Hertel04a,*Porrati05,*Niedoba05,*Vila09} in bubble materials, \cite{Masseboeuf09,*Jourdan09} in disks--shaped \cite{Thiaville03,Hertel06} and astroid--shaped nanodots\cite{Xing08}. The ultrafast switching of the vortex core magnetization open doors to consider the vortex state nanoparticles as promising candidates for magnetic elements of storage devices. There are different scenarios of the switching process: (i) The symmetric or so--called punch--through core reversal takes place under the action of DC magnetic field applied perpendicularly to the magnet plane.\cite{Okuno02,Thiaville03,Kravchuk07a,Vila09} This reversal process as a rule is mediated by creation of two Bloch points.\cite{Thiaville03} However single Bloch point scenario was also mentioned in \citet{Thiaville03}. (ii) The switching under the action of different in--plane AC magnetic fields or by a spin polarized currents, \cite{Waeyenberge06,Xiao06,Hertel07,Yamada07,Kravchuk07c,Sheka07b} is accompanied by the temporary creation and annihilation of the vortex--antivortex pair. The latter is accompanied by Bloch point creation \cite{Hertel06}.

The purpose of the current work is to study the magnetization structure of the Bloch point of the  spherical nanosized particle. As opposed to bubble films, where the static Bloch point results from the transition between Bloch lines, \cite{Malozemoff79,Hubert98} and vortex nanodots, where the Bloch point dynamically appears during the vortex core switching process, \cite{Thiaville03,Waeyenberge06} the Bloch point in the nanosphere is an example of ``pure'' singularity without surrounding. Such a singularity is in some respect the only stable singularity in ferromagnet.\cite{Thiaville03} We consider different types of Bloch point and classify them in terms of vortex parameters. The conventional magnetization distribution in the Bloch point is generalized for the radial--dependent one. Such radial distribution becomes important for the Bloch point nanosphere under the action of nonhomogeneous magnetic field. We show that radial gradient field can stabilize the Bloch point and compute the magnetization structure, which is in a good agrement with performed spin--lattice simulations.

The paper is organized as follows. In Sec.~\ref{sec:model} we describe the model and present the classification of different Bloch point types (Sec.~\ref{sec:class}). The energetic analysis and the Bloch structure is analyzed in Sec.~\ref{sec:structure}. In order to stabilize the Bloch point inside the nanosphere, we consider the the influence of external gradient field on the magnetization structure. The Bloch point solution becomes radially dependent: we calculate the magnetization structure analytically in Sec.~\ref{sec:field}. In Sec.~\ref{sec:simulations} we study the Bloch point structure numerically, in particular, the problem of stability. We discuss our results in Sec.~\ref{sec:conclusion}. In Appendix \ref{sec:weak} we analyze the Bloch point structure under the influence of weak fields using the linearized equations.

\section{The model and the Bloch point solutions}
\label{sec:model}

Let us consider the classical isotropic ferromagnetic sphere of the radius $R$. The continuum dynamics of the magnetization can be described in terms of the magnetization unit vector $\vec{m} = \vec{M}/M_\text{S} =\left(\sin\Theta\cos\Phi,\sin\Theta\sin\Phi,\cos\Theta\right)$, where $\Theta$ and $\Phi$ are, in general, functions of the coordinates and the time, and $M_S$ is the saturation magnetization. The total energy $E$ of such a sphere, normalized by $4\pi M_S^2 V$ with $V=\frac43\pi R^3$ reads:
\begin{subequations} \label{eq:Energies}
\begin{equation} \label{eq:Energy}
\mathscr{E} = \mathscr{E}^{\text{ex}} + \mathscr{E}^{\text{f}} + \mathscr{E}^{\text{ms}}.
\end{equation}
The first term in \eqref{eq:Energy} is dimensionless exchange energy:
\begin{equation} \label{eq:E-ex}
\mathscr{E}^{\text{ex}} = \frac{3}{8\pi}\varepsilon\int \mathrm{d}\vec{r} \left[ \left(\vec{\nabla}\Theta\right)^2 + \sin^2\Theta\left(\vec{\nabla}\Phi\right)^2\right]
\end{equation}
with $\varepsilon=\ell^2/R^2$ being reduced exchange length, $\ell=\sqrt{A/4\pi M_S^2}$ being the exchange length, $A$ being the exchange constant and $\vec{r}=(x,y,z)/R$ being the reduced radius--vector. The second term determines the interaction with external magnetic field $\vec{H}$:
\begin{equation} \label{eq:E-f}
\mathscr{E}^{\text{f}} = -\frac{3}{4\pi}\int \mathrm{d}\vec{r} \left(\vec{m}\cdot \vec{h} \right),
\end{equation}
where $\vec{h}=\vec{H}/4\pi M_S$ is a reduced external field. We will discuss the influence of external field later, see Sec.~\ref{sec:field}. The last term determines the reduced magnetostatic energy:
\begin{equation} \label{eq:E-ms}
\mathscr{E}^{\text{ms}} = -\frac{3}{8\pi}\int \mathrm{d}\vec{r} \left(\vec{m}\cdot \vec{h}^{\text{ms}} \right),
\end{equation}
\end{subequations}
where $\vec{h}^{\text{ms}}=\vec{H}^{\text{ms}}/4\pi M_S$ is a reduced magnetostatic field $\vec{H}^{\text{ms}}$. Magnetostatic field $\vec{h}^{\text{ms}}$ satisfies the Maxwell magnetostatic equations \cite{Hubert98,Stohr06}
\begin{equation} \label{eq:Maxwell}
\begin{cases}
\vec{\nabla}\times \vec{h}^{\text{ms}} = 0,\\
\vec{\nabla}\cdot \vec{h}^{\text{ms}} = 4\pi\lambda,
\end{cases}
\end{equation}
which can be solved using magnetostatic potential, $\vec{h}^{\text{ms}} = - \vec{\nabla} \psi$. The source of the field $\vec{h}^{\text{ms}}$ are magnetostatic charges: volume charges $\lambda \equiv - (\vec{\nabla}\cdot \vec{m})/4\pi$ and surface ones $\sigma \equiv  (\vec{m}\cdot\vec{n})/4\pi$ with $\vec n$ being the external normal. The magnetostatic potential inside the sample reads:
\begin{subequations}
\begin{align} \label{eq:MS-psi1}
&\psi(\vec{r}) = \int_V \mathrm{d}\vec{r}' \frac{\lambda(\vec{r'})}{|\vec{r}-\vec{r'}|} + \int_S \mathrm{d} S' \frac{\sigma(\vec{r'})}{|\vec{r}-\vec{r'}|}\\
\label{eq:MS-psi2}
&\equiv \frac{1}{4\pi} \int_V\mathrm d\vec r' \Bigl(\vec m(\vec r')\cdot \nabla_{\vec r'}\Bigr) \frac{1}{|\vec{r}-\vec{r'}|}.
\end{align}
\end{subequations}

The equilibrium magnetization configuration is determined by minimization of the energy functional \eqref{eq:Energies}, which leads to the following set of equations:
\begin{equation} \label{eq:str}
\varepsilon\nabla^2\vec{m}=\vec{\nabla}\psi, \quad \nabla^2\psi = \vec{\nabla}\cdot \vec{m}.
\end{equation}

\subsection{Classification of singularities}
\label{sec:class}

Let us start the Bloch point as a particular solution of \eqref{eq:str}. In the exchange approach the simplest hedgehog--type Bloch point  is characterized by the magnetization distribution of the form $\vec{m} = \vec{r}/r$  with a singularity at the origin. Using spherical frame of reference for the radius--vector $\vec{r}$ with the polar angle $\vartheta$ and azimuthal one $\varphi$, one can describe the magnetization angles of such a Bloch point as follows: $\Theta=\vartheta$ and $\Phi=\varphi$. The energy of the Bloch point in the exchange approach reads\cite{Doering68}
\begin{equation} \label{eq:Energy-BP}
\mathscr{E}_0^{\text{ex}} =3\varepsilon, \qquad E_0^{\text{ex}} = 4\pi A R.
\end{equation}
This interaction is invariant with respect to the joint rotation of all magnetization vectors, which gives a possibility to consider family of solutions with different rotation angles.\cite{Feldtkeller65b,Doering68}

We consider the following singular magnetization distribution:
\begin{equation} \label{eq:tBP-str}
\Theta(\vartheta) = p\vartheta + \pi(1-p)/2, \; \Phi(\varphi) = q\varphi+\gamma, \quad p, q = \pm1,
\end{equation}
which describes a three--parameter Bloch point. We refer to the parameter $q=\pm1$ as to vorticity of the Bloch point and $p=\pm1$ as to its polarity using the conventional symbols for magnetic vortices. The last parameter  $\gamma$ describes the azimuthal rotational angle of the Bloch point.\cite{Feldtkeller65b,Doering68}

\begin{figure*}
\begin{center}
\subfloat[$p=q=1$]{\label{fig:BP11}\includegraphics[width=0.2\linewidth]{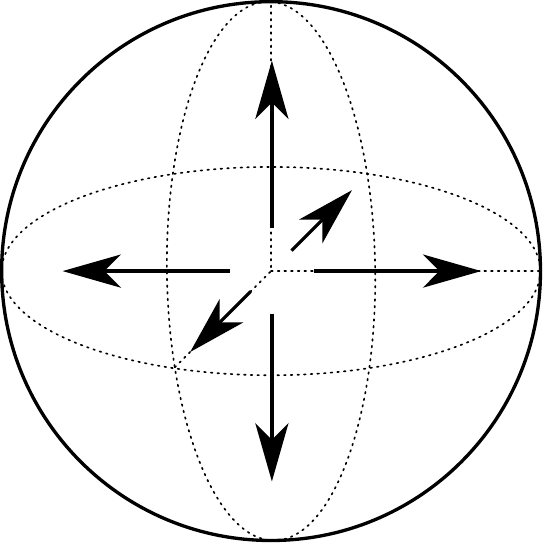}}\hfill
\subfloat[$p=-1\;q=1$]{\label{fig:BPm11}\includegraphics[width=0.2\linewidth]{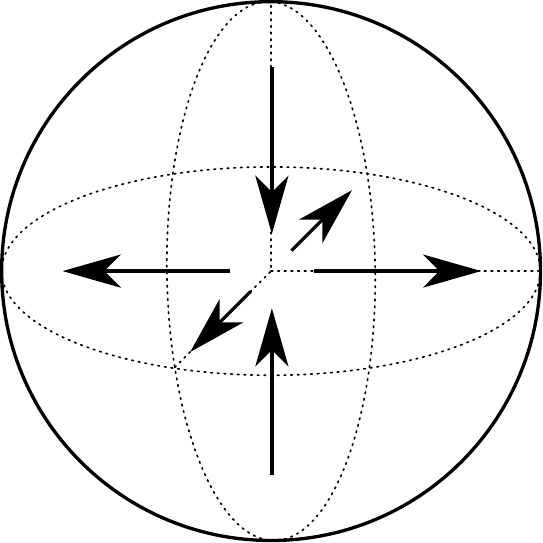}}\hfill
\subfloat[Vortex switching]{\label{fig:switchVortex}\includegraphics[width=0.084\linewidth]{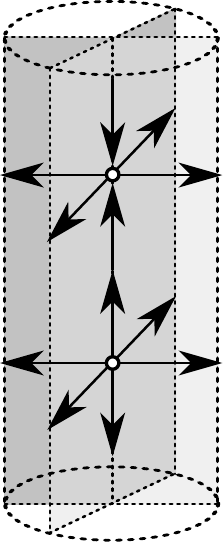}}\hfill
\subfloat[$p=1\;q=-1$]{\label{fig:BP1m1}\includegraphics[width=0.2\linewidth]{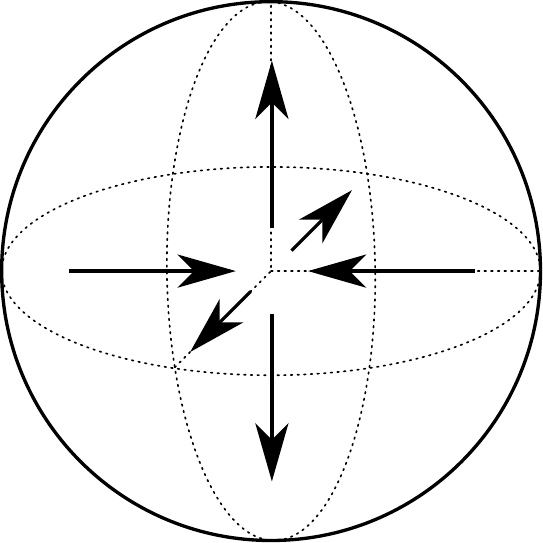}}\hfill
\subfloat[$p=q=-1$]{\label{fig:BPm1m1}\includegraphics[width=0.2\linewidth]{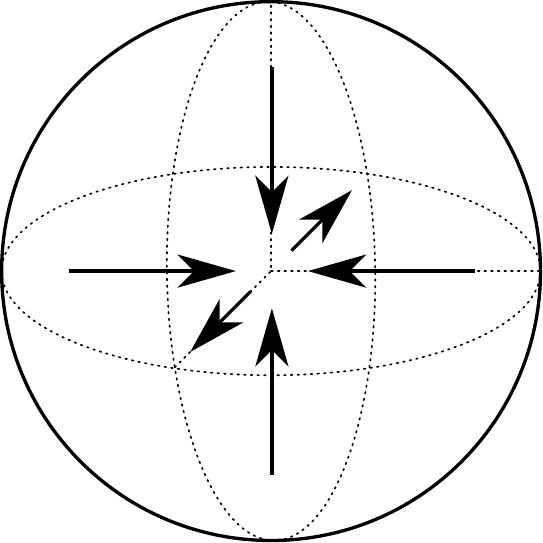}}\hfill
\subfloat[Antivortex switching]{\label{fig:switchAntivortex}\includegraphics[width=0.084\linewidth]{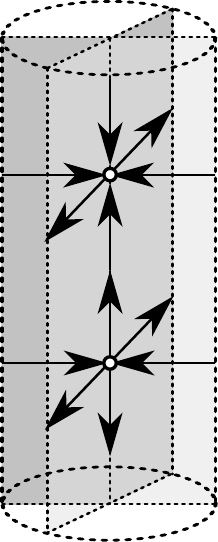}}
\end{center}
\caption{Schematic of different types of Bloch points. Magnetization distribution in azimuthal vortex Bloch points in sphere, see Figs. \ref{fig:BP11}, \ref{fig:BPm11}, and both Bloch points in axial part of cylinder-shaped sample during the vortex polarity switching  process, see Fig.~\ref{fig:switchVortex}. The same is for azimuthal antivortex Bloch points, see Figs.~\ref{fig:BP1m1}, Fig.~\ref{fig:BPm1m1}, and both singularities in axial part of astroid--shaped sample during the switching, see Fig.~\ref{fig:switchAntivortex}.}
\label{fig:schematic}
\end{figure*}

We refer to the micromagnetic singularity \eqref{eq:tBP-str} as to BP$^p_q$. For example, the hedgehog--type Bloch point is a vortex Bloch point with positive polarity ($p=1$, $q=1$, $\gamma=0$). The schematic of magnetization distribution in different types of Bloch points is presented on Fig.~\ref{fig:schematic}. The analogy between Bloch point and vortices comes from the symmetric or punch--through vortex polarity switching process under the action of DC perpendicular magnetic field.\cite{Thiaville03} This reversal process as a rule is mediated by creation of two Bloch points.\cite{Thiaville03} For example, two singularities, BP$_1^1$ and BP$_1^{-1}$ describe intermediate state between two vortices with opposite polarities, see Fig.~\ref{fig:switchVortex}. It is also instructive to mention that a single Bloch point can be imagined as a composite of two vortices with opposite polarities: such a singularity can appear in 3D Euclidean space during the vortex polarity switching process in antiferromagnets \cite{Galkina95,Senthil04}. All four distributions for different signs of $p$ and $q$ can be observed during symmetrical Bloch points injection in polarity switching process of vortices\cite{Thiaville03} and antivortices\cite{Xing08}.

Topological properties of the Bloch point can be described by the topological (Pontryagin) index
\begin{equation} \label{eq:Pontryagin}
Q = \frac{1}{4\pi} \int\sin\Theta(\vec r)\mathrm d\Theta(\vec r)\mathrm d\Phi(\vec r) = pq.
\end{equation}
Different Bloch point distributions with equal $Q$ are topologically equivalent: e.g., BP$^{-1}_{-1}$ can be obtained from BP$^1_1$ by simultaneous rotation of all magnetization vectors by $\pi$ in vertical plane, and BP$^1_{-1}$ transforms to BP$^{-1}_1$ by rotation by $\pi/2$ in vertical plane. Note that similar topological notations were introduced by \citet{Malozemoff79} for magnetic bubbles.\footnote{Bloch points in magnetic bubbles were classified by \citet{Malozemoff79} using the flux $N$ of gyrotropic vector. Simple calculations show that the value of such a flux is opposite to the topological density \eqref{eq:Pontryagin}, $N =-Q$.}

\subsection{Magnetization structure of Bloch points}
\label{sec:structure}

The most strong exchange interaction is invariant with respect to the rotation angle $\gamma$. Such degeneracy is removed under account of magnetostatic interaction. It is worth noting that the problem of stray field influence on the Bloch point energetics has a long story. Feldtkeller in his pioneer work\cite{Feldtkeller65b} used a so--called pole avoidance principle, see e.g. Ref.~\onlinecite{Aharoni96}: the magnetostatic tries to avoid any sort of volume or surface charge. In this way he calculated the angle $\gamma$ from the condition that the total volume magnetostatic charge $\int \lambda(\vec r)  \mathrm{d}\vec{r} =0$, where $\lambda(\vec{r})$ is the charge density. For the Bloch point given by Ansatz \eqref{eq:tBP-str} it has a form $\lambda(\vec r) = -\left[p \sin^2\vartheta + \cos\gamma(\cos^2\vartheta+1)\right]/4\pi r$ and leads to the rotation angle
\begin{equation} \label{eq:E-ms-F}
\gamma_\text{F} = \arccos\left(-\frac{p}{2}\right) =
\begin{cases}
120^\circ\!\!, \!\!& \!\! p=+1, \\
60^\circ\!\!,  \!\!& \!\! p=-1.
\end{cases}
\end{equation}
In it interesting to note that the same value $\gamma_\text{F}$ also corresponds to absence of the total surface charge, $\int \sigma(\vec r)  \mathrm{d}S =0$, where the surface charge density $\sigma(\vec r) = \left(p\cos^2\vartheta + \cos\gamma\sin^2\vartheta\right)/4\pi$.

Another approach was put forward by \citet{Doering68}, who determined the equilibrium angle of $\gamma$ by minimizing the energy
\begin{equation} \label{eq:E-ms-1}
\mathscr E^{\text{ms}}_{\text{D}} = \frac{3}{8\pi} \int_V \mathrm d\vec r (h^{\text{ms}})^2
\end{equation}
and obtained
\begin{equation} \label{eq:E-ms-D}
\gamma_\text{D} = \arccos\left(-\frac{11}{29}\right) \approx 112.3^\circ.
\end{equation}
However one has to emphasize that the equilibrium angle \eqref{eq:E-ms-D} minimizes only the inner part of the magnetostatic energy because   the integration in \eqref{eq:E-ms-1} is carried  over the sample volume $V$ while  the outer part of stray field is ignored. Note the similar approach was used in quite recent paper,\cite{Elias11} where a magnetization contraction was taken into account.

The aim of this section is to find the equilibrium rotation angle which minimizes the total magnetostatic energy.
In order to derive the magnetostatic energy of Bloch points \eqref{eq:tBP-str}, we calculate first magnetostatic potential \eqref{eq:MS-psi2} using an expansion of $1/|\vec{r}-\vec{r}'|$ over the spherical harmonics,
\begin{equation*} \label{eq:expansion-on-Ylm}
\frac{1}{|\vec{r} - \vec{r}'|}  = \frac{1}{r_>}\sum_{l=0}^\infty\! \sum_{m=-l}^l \frac{4\pi}{2l+1} \!\!\left(\frac{r_<}{r_>}\right)^{\!\!l} \!\! Y_{lm}\left(\vartheta,\varphi\right) Y_{lm}^\star\left(\vartheta',\varphi'\right)
\end{equation*}
with $r_<=\min(r,r')$ and $r_>=\max(r,r')$
which results in
\begin{subequations} \label{eq:psi4p}
\begin{align*}
\psi_{q=1}^p(\vec{r}) &= p\pi r + \frac{\pi}{3}(9r-8)\cos\gamma+\pi r(p-\cos\gamma)\cos^2\vartheta,\\
\psi_{q=-1}^p(\vec{r}) &= p\pi r(1+\cos^2\vartheta)+\pi r \cos(2\varphi+\gamma)\sin^2\vartheta.
\end{align*}
\end{subequations}
Simple calculations show that the magnetostatic energy of the antivortex Bloch point does not depend on $\gamma$ and $\mathscr{E}^\text{ms}_{q=-1} = {7}/{30} \approx 0.23$. In contrast to this, the vortex Bloch point energy depends on the rotation angle $\gamma$ and has the form
\begin{equation} \label{eq:Ep1}
\mathscr{E}^{\text{ms}\;p}_{q=1}(\gamma) = \frac{1}{30}\left(7+4 p \cos\gamma+ 4\cos 2 \gamma \right).
\end{equation}
The equilibrium value of rotation angle $\gamma_0$ corresponds to the minimum of the energy \eqref{eq:Ep1}. It gives
\begin{equation} \label{eq:gamma0Ems1}
\gamma_0 = \arccos\left(-\frac{p}{4}\right)\approx
\begin{cases}
105^\circ\!\!, \!\!& \!\! p=+1, \\
76^\circ\!\!,  \!\!& \!\! p=-1.
\end{cases}
\end{equation}

Let us compare Bloch point energies \eqref{eq:Ep1} for above mentioned approaches: the energy of \citet{Feldtkeller65b} Bloch point $\mathscr{E}^{\text{ms}\;p}_{q=1}(\gamma_\text{F}) = 0.1$, for \citet{Doering68} Bloch point  one has $\mathscr{E}^{\text{ms}}(\gamma_\text{D}) \approx 0.088$, the result by \citet{Elias11} is $\mathscr{E}^{\text{ms}}_1(\gamma_\text{EV}) \approx 0.089$. The minimal energy has a Bloch point with the rotation angle $\gamma_0$, see \eqref{eq:gamma0Ems1}:
\begin{equation}\label{eq:Ems(gamma-0)}
\mathscr{E}^{\text{ms}\;p}_{q=1}(\gamma_0) = \frac{1}{12} \approx 0.083.
\end{equation}

\begin{figure}
\includegraphics[width=\columnwidth]{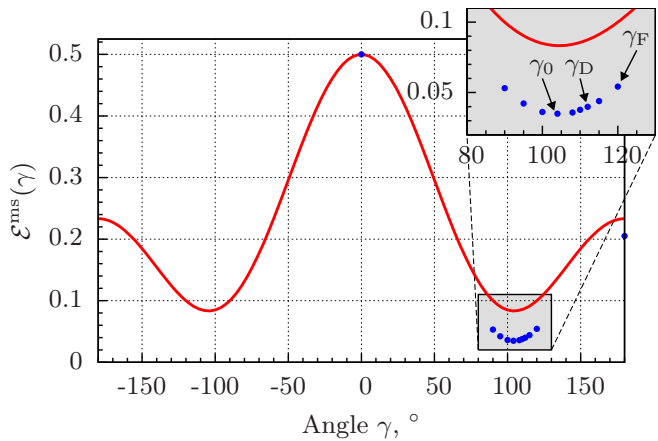}
\caption{(Color online.) The Bloch point energy \emph{vs} rotation angle for BP$^1_1$: analytical result \eqref{eq:Ep1} (solid curve) and simulations (symbols). Simulations parameters: sphere diameter $2R = 35a_0$, exchange length $\ell = 3.95a_0$, damping parameter $\eta=0.5$.
}
\label{fig:Ems}
\end{figure}

In order to verify our results  we performed numerical spin--lattice simulations, see details in Sec.~\ref{sec:simulations}. We compare analytical dependence $\mathscr{E}^{\text{ms}\;p=1}_{q=1}(\gamma)$, see Eq.~\eqref{eq:Ep1}, with the discrete energy \eqref{eq:Hamiltonian}, extracted from simulations, see Fig.~\ref{fig:Ems}. Both dependencies are matched in maximum at $\gamma = 0$. Comparison can be provided by calculating energy gain $\Delta \mathscr E(\gamma) = \mathscr{E}^\text{ms}_\text{max} - \mathscr{E}^\text{ms}(\gamma)$ for different rotation angles $\gamma$. According to simulation results the energy gain for mentioned above angles read:
\begin{equation*}
\Delta \mathscr E(\gamma_\text{F}) \approx 0.446,\quad
\Delta \mathscr E(\gamma_\text{D}) \approx 0.460,\quad
\Delta \mathscr E(\gamma_0) \approx 0.465.
\end{equation*}
The maximum energy gain takes place for $\gamma_0$, which corresponds to the energy minimum in a good agrement with our analytical result \eqref{eq:gamma0Ems1}.

\section{The Bloch points in external field}
\label{sec:field}

The Bloch point does not form a ground state of a magnetic sphere. It corresponds to the saddle point (sphaleron) of the energy functional \cite{Manton04}. This brings up the question: How to stabilize the Bloch point? In this section we show that one way to achieve this goal is to  apply a magnetic field which has the same symmetry as the hedgehog Bloch point with $\vec m = \vec{r}/r$, \emph{i.~e.} a radial symmetric magnetic gradient magnetic field in the form
\begin{equation} \label{eq:h-grad}
\vec h = b \vec r\;.
\end{equation}

Under the action of the space dependent magnetic field \eqref{eq:h-grad} the magnetization distribution also becomes space dependent. We take into account possible dependence by the following radial Bloch point Ansatz
\begin{equation} \label{eq:tBP-gamma(r)}
\Theta(\vartheta) = p\vartheta + \pi(1-p)/2, \; \Phi(r,\varphi) = q\varphi+\gamma(r)
\end{equation}
with a radially dependent parameter $\gamma(r)$ in comparison with Eq.~\eqref{eq:tBP-str}. The form of this Ansatz will justified by numerical simulations in Sec.~\ref{sec:simulations}.

Inserting Eq. \eqref{eq:tBP-gamma(r)} into Eq. \eqref{eq:E-ex} for the exchange energy of such magnetization distribution we get
\begin{subequations} \label{eq:BPen}
\begin{equation} \label{eq:BPenEx}
\mathscr{E}^{\text{ex}} = 3\varepsilon + \varepsilon\int_0^1 \left(\frac{ \mathrm d\gamma}{\mathrm d r}\right)^2 r^2 \mathrm{d}r.
\end{equation}
The magnetostatical potential of the Bloch point \eqref{eq:tBP-gamma(r)} reads
\begin{align*}
\psi_{q=1}^{p=1}(\vec{r}) &= -\frac{4\pi}{3}\int_r^1 \left[ 1+2\cos\gamma(r')\right]\mathrm dr'-\\
&-\frac{4\pi}{3} \frac{3\cos^2\vartheta-1}{r^3}\int_0^r r'^3 \left[\cos\gamma(r')-1\right]\mathrm dr'.
\end{align*}
Here and below we consider the case of BP$_1^1$ only. The magnetostatic energy of such a Bloch point has the form
\begin{equation} \label{eq:BPenMs}
\mathscr E^\text{ms} = \frac{1}{10}\,\int\limits_0^1 r^2\left[7+4\,\cos\gamma(r)+4\,\cos 2\gamma(r)\right] \mathrm dr.
\end{equation}
From Eq. \eqref{eq:E-f} we obtain that the Bloch point interaction with magnetic field can be  expressed as follows
\begin{equation} \label{eq:BPenF}
\mathscr E^\text{f}=-2b\int\limits_0^1 r^3\cos\gamma(r) \mathrm dr.
\end{equation}
\end{subequations}
 By minimizing the total energy, ${\delta \mathscr{E}}/{\delta \gamma} = 0$,
we obtain that the equilibrium distribution $\gamma(r)$ is a solution of the  following nonlinear differential equation
\begin{equation} \label{eq:varEqRad}
\varepsilon\frac{ \mathrm d^2\gamma}{\mathrm d r^2} + \frac{2\varepsilon}{r}\frac{ \mathrm d\gamma}{\mathrm d r} + \frac{1}{5}\sin\gamma + \frac{2}{5}\sin2\gamma - br\sin\gamma = 0
\end{equation}
augmented by boundary conditions of the form
\begin{equation} \label{eq:varEqRad_BC}
\frac{\mathrm{d}\gamma}{\mathrm{d}r}\biggr|_{r=0} = \frac{\mathrm{d}\gamma}{\mathrm{d}r}\biggr|_{r=1} = 0.
\end{equation}

\begin{figure}
\includegraphics[width=\columnwidth]{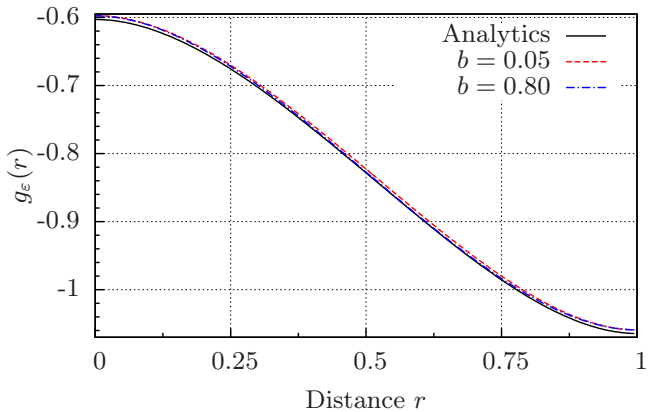}
\caption{(Color online.) Reduced rotation angle $g_\varepsilon(r)$, see \eqref{eq:smallFieldsEstimation} for different field intensities and $\varepsilon = 0.05$: analytical result \eqref{eq:f(xi)} (solid curve) and numerical solution of \eqref{eq:varEqRad} (dashed curves).}
\label{fig:fbyr-eps}
\end{figure}
In the case of weak fields one can linearize Eq. \eqref{eq:varEqRad} in the vicinity of spatially uniform solution \eqref{eq:gamma0Ems1} and obtain that
\begin{equation} \label{eq:smallFieldsEstimation}
\gamma(r) \approx \gamma_0 + b g_\varepsilon(r), \qquad |b|\ll 1.
\end{equation}
An explicit form of the function $g_\varepsilon(r)$ is calculated in Appendix \ref{sec:weak}.
The comparison with numerical solution of Eq.~\eqref{eq:varEqRad} shows a quite good agreement up to relatively strong fields ($b\lesssim1$), see Fig.~\ref{fig:fbyr-eps}.

\begin{figure*}
\subfloat[Critical behaviour]{\label{fig:criticalBehaviour}\includegraphics[width=0.495\linewidth]{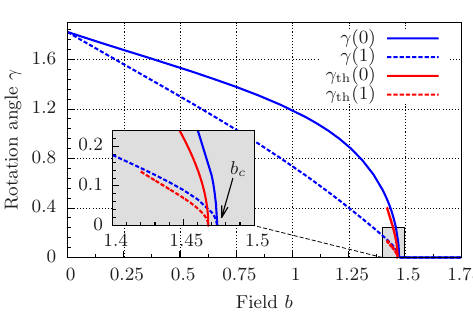}}\hfill
\subfloat[Phase diagram]{\label{fig:phaseEpsVsB}\includegraphics[width=0.495\linewidth]{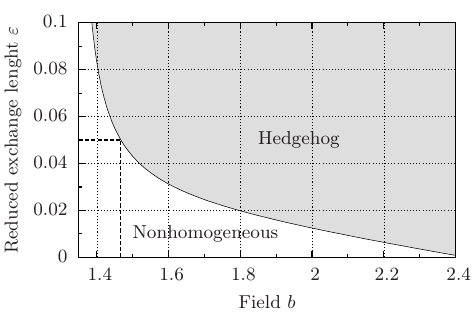}}
\caption{(Color online.) Bloch point under the action of the gradient field.
(a):
Rotation angle \emph{vs} field intensity $b$ near the critical field $b_c \approx 1.473$ from numerical solution of Eq.~\eqref{eq:varEqRad} (blue curves) and theoretical estimation by Eq.~\eqref{eq:gamma-cri} (red curves) with $\varepsilon=0.05$. Solid lines correspond to the rotation angle $\gamma(0)$ and dashed line to $\gamma(1)$.
(b): Phase diagram for solutions of Eq.~\eqref{eq:varEqRad}. The upper (hedgehog) phase correspond to the solution $\gamma=0$, the lower (nonhomogeneous) one to the radial--dependent Bloch point with $\gamma(r)$. Dashed lines correspond to analytical result for the critical field $b_c = 1.465$ for $\varepsilon = 0.05$, see text.}
\label{fig:hcr}
\end{figure*}
Another limiting case is realized in the case of strong magnetic fields when the Bloch point magnetization is parallel to the external field. In this case the rotation angle is  $\gamma=0$ (mod $\pi$).

To describe the behavior of the Bloch point in a critical region $b\approx b_c$ where the spatially non-uniform distribution transforms to the  spatially uniform one,  we use a variational approach with a  two--harmonics trial function $\gamma(\vec r)\approx \alpha_0 + \alpha_1 \cos\pi r$. Near the critical point  $\alpha_0, \alpha_1\ll1$. We expand the total energy in a Taylor series up to the fourth order with  respect to $\alpha_0$ and to the second order with respect to $\alpha_1$. By excluding $\alpha_1$ and keeping terms not higher than $\alpha_0^4$, we get
\begin{equation} \label{eq:E-cri}
\mathscr{E}(\gamma) \approx \mathscr{E}_0 + p_2(b,\varepsilon) \alpha_0^2 + p_4(b,\varepsilon)\alpha_0^4,
\end{equation}
The energy \eqref{eq:E-cri} as a function of $\alpha_0$ has a double--well shape ($p_2(b,\varepsilon)<0$)  for $b<b_c$ with  the critical magnetic field $b_c$ given by
\begin{equation}\label{eq:cr_b}
b_c(\varepsilon) \approx 1.8 -21.6 \varepsilon +\sqrt{0.4 -20.2 \varepsilon +467 \varepsilon ^2}\end{equation}
In the critical region when $0<b_c(\varepsilon)-b\ll b_c(\varepsilon)\;,$
\begin{equation} \label{eq:gamma-cri}
\begin{split}
\alpha_0(b)&\approx a(\varepsilon) \sqrt{b_c(\varepsilon)-b}
\end{split}
\end{equation}
For $b>b_c,~~p_2>0$ and the  function Eq. \eqref{eq:E-cri} has a minimum for  $\alpha_0=0$. It corresponds to $\gamma=0$. Numerical integration of Eq.~\eqref{eq:varEqRad} for $\varepsilon=0.05$ shows that the phase transition occurs when $b_c \approx 1.473$, see Fig.~\ref{fig:hcr}. It agrees well with the value $b_c(0.05) \approx 1.465$ obtained from Eq. \eqref{eq:cr_b}. The critical behavior predicted by \eqref{eq:gamma-cri} is also confirmed by our numerical simulations (see Fig. \ref{fig:criticalBehaviour}).

\section{Numerical study of the Bloch Point structure}
\label{sec:simulations}

\begin{figure}
\includegraphics[width=\columnwidth]{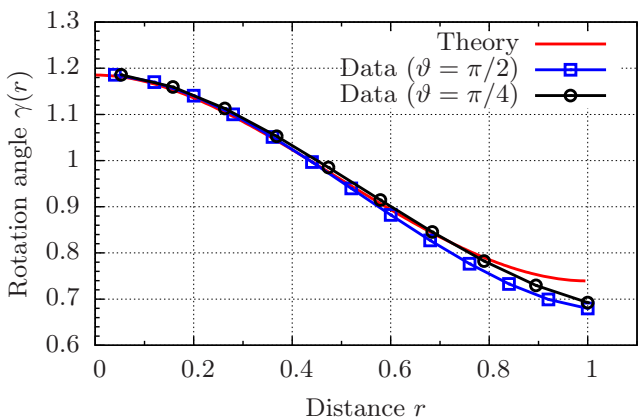}
\caption{(Color online.) Radial dependence of rotation angle $\gamma$ in spherical particle. Line: numerical integration of \eqref{eq:varEqRad}. Symbols: \texttt{SLaSi} simulations for crystallographic directions [110] and [111]. Parameters are the same as in Fig.~\ref{fig:Ems}.}
\label{fig:sphericalLayer}
\end{figure}

In order to check analytical results about Bloch point structure, we performed simulations using in--house developed spin--lattice simulator \texttt{SLaSi}\cite{SLaSi} that solves Landau--Lifshitz--Gilbert equation in terms of spins
\begin{equation*}
\frac{\mathrm d\vec S_n}{\mathrm dt} = - \frac{1}{\hbar} \left[ \vec S_n \times \frac{\partial \mathscr H}{\partial \vec S_n} \right] - \frac{\eta}{S} \left[ \vec S_n \times \frac{\mathrm  d \vec S_n}{\mathrm d t} \right],
\end{equation*}
where $\mathscr H$ is a lattice Hamiltonian of the classical ferromagnet:
\begin{equation} \label{eq:Hamiltonian}
\begin{split}
&\mathscr{H} = -\frac{J}{2} \sum_{\left(n,\delta\right)} \vec{S}_{n}\cdot \vec{S}_{n+\delta}  + 2\mu_\text{B} \vec H \sum_{n} \vec S_{n} \\
&+ 2\mu_\text{B}^2 \sum_{n\neq k}\left[ \frac{(\vec S_{n}\cdot \vec S_{k})}{{r_{nk}}^3} - 3 \frac{(\vec S_{n} \cdot \vec r_{nk})(\vec S_{k}\cdot \vec r_{nk})}{{r_{nk}}^5} \right].
\end{split}
\end{equation}
Here  $\vec{S}_{n}$ is a classical spin vector with fixed length $S$ in units of action on the site $n$ of a three--dimensional cubic lattice with lattice constant $a_0$, $J$ is the exchange integral, $\mu_\text{B}$ is Bohr magneton, $\vec r_{nk}$ is the radius--vector between $n$-th and $k$-th nodes, $\eta$ is a damping parameter, $\vec H$ is external magnetic field and $\delta$ runs over six nearest neighbors. Integration is performed by modified 4--5~order Runge--Kutta--Fehlberg method (RKF45) and free spins on the surface of the sample. \footnote{We take the integration time step $\Delta t = 0.025 \omega_0^{-1}$ for the accuracy $\Delta S_\text{max} = \max\limits_{i, n} |S_{i, n}^4-S_{i, n}^5| < 0.01$, where $\omega_0 = JS^2/\hbar$, $i=x,y,z$ and superscript for $S_{i, n}$ indicates integration order. We used modified RKF45 scheme, where the time step is changed only if accuracy goal for $\Delta S_\text{max}$ is not reached for avoiding of noise from step changing.}

Numerically we checked the Bloch point structure, given by the radial--dependent Ansatz \eqref{eq:tBP-gamma(r)}. by modeling spherically--shaped sample with diameter $2R = 35a_0$ (such a sample consists of 24\,464 nodes with nonzero spin), and exchange length $\ell = 3.95a_0$ ($\varepsilon = 0.05$). In order to stabilize the Bloch point we applied the gradient magnetic field with $b = 1.0$. By modeling  the overdamped dynamics we observed that the Bloch point structure quickly relaxes to the state similar to one, given by \eqref{eq:tBP-gamma(r)}: The polar Bloch point angle $\Theta(\vec r)$ does not deviate from $\vartheta$ within the accuracy $ 0.099$. The azimuthal angle is well also well--described by \eqref{eq:tBP-gamma(r)} with the radial--dependent rotation angle $\gamma(r)$, see Fig.~\ref{fig:sphericalLayer}. Simulations were performed for crystallographic directions [111] ($\vartheta=\pi/4$) and [110] ($\vartheta\approx \pi/2$, the plane is  shifted by $z=-0.5a_0$ from the origin). One can see from Fig.~\ref{fig:sphericalLayer} that numerical data are well confirmed by analytical curve $\gamma(r)$, calculated as numerical solution of \eqref{eq:varEqRad}.

\begin{figure*}
\includegraphics[width=\linewidth]{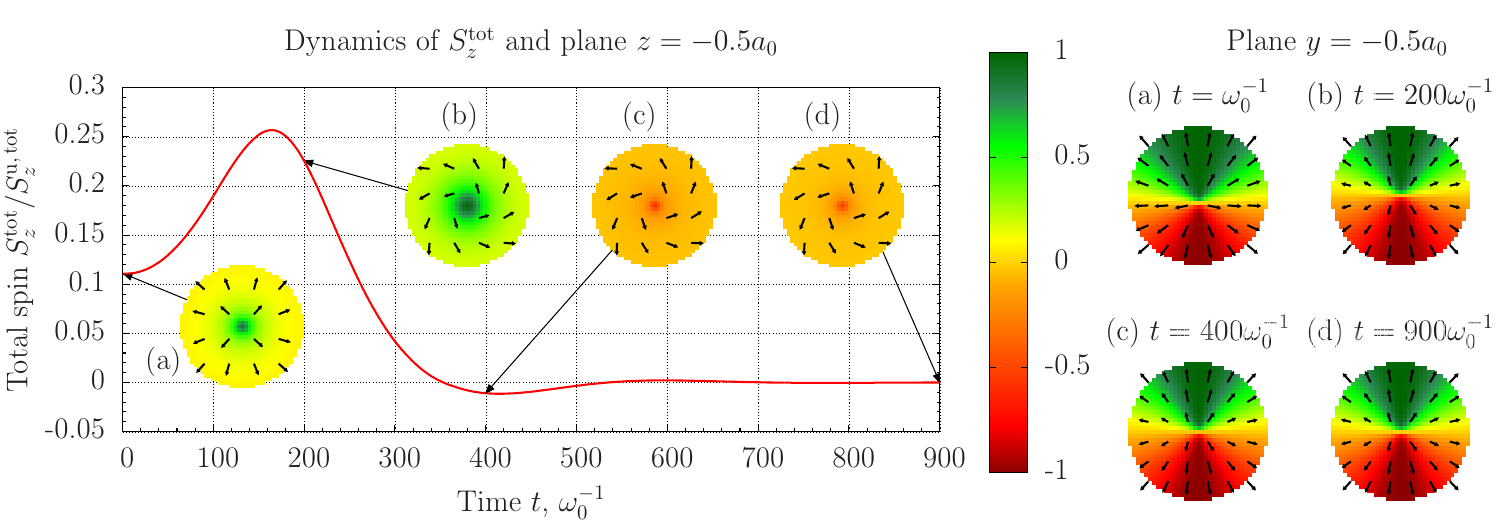}
\caption{(Color online.) Dynamics of total spin along $z$--axis of the sample. The Bloch point is initially shifted by $\Delta z = -2a_0$ from center of the sample. Insets show magnetization distribution in $z = -0.5a_0$ and $y = -0.5a_0$ planes in different times. Color bar indicates $S_{z,\vec n}$ for different lattice nodes. Applied field amplitude $b = 1$, other parameters are the~same~as~in~Fig.~\ref{fig:Ems}.}
\label{fig:simulation}
\end{figure*}

To validate our theory we performed also direct stability check. Numerically we check the stability of the Bloch point against the shift of its position. We start simulations with the Bloch point state using Ansatz--function \eqref{eq:tBP-gamma(r)}, which is shifted along $\hat{z}$--axis by $\Delta z= -2a_0$. We also apply $\gamma(r,t=0) = 3^\circ$ in order to break the symmetry. For rapid relaxation we used in most of simulations the overdamped regime (the damping parameter $\eta = 0.5$). We checked the shift of the Bloch point by controlling the total spin projections: only for the Bloch point, situated at the sample origin, the total spin $S_x^{\text{tot}} = S_y^{\text{tot}} = S_z^{\text{tot}} = 0$.

The temporal evolution of initially shifted Bloch point is presented in Fig.~\ref{fig:simulation} for the Bloch point sample with $2R = 35a_0$ (24\,456 nodes) in applied field with $b = 1$, see also the supplementary video \footnote{supplementary video \url{http://slasi.rpd.univ.kiev.ua/sm/pylypovskyi12}}. Originally the Bloch point was shifted down from the origin which corresponds to $S_z^\text{tot} > 0$, see inset (a). During the evolution a number of magnons are generated, inset (b). After quick damping of oscillations, the micromagnetic singularity goes to the sample origin, see inset (c). The relaxation process consists of two parts: (i) The rotation angle $\gamma(r)$ changes its value from initial uniform one to the final nonhomogeneous state during a time $\tau_\gamma \approx 500\omega_0^{-1}$. (ii) The relaxation of $S_z^\text{tot}$ component of total spin of the sample tooks approximately the same time. During all simulations time $\left|S_x^\text{tot}\right| \approx \left|S_y^\text{tot}\right|\lesssim 10^{-11}$.

\section{Conclusion}
\label{sec:conclusion}

To summarize, we study the magnetization structure of the Bloch point. In spite of the fact that the Bloch point as a simplest 3D topological singularity was studied during a long time, from the pioneer papers by \citet{Feldtkeller65b} and \citet{Doering68}, see also for review Refs.~\onlinecite{Malozemoff79,Hubert98}, the problem of the Bloch point structure still causes discussions.\cite{Thiaville03,Khodenkov10,Elias11} The point is that the most strong exchange interaction determines only the relative magnetization distribution accurate within the rotation angle $\gamma$. This rotation angle, which is determined by the magnetostatic interaction, is most questionable: its value is equal to $120^\circ$ according to \citet{Feldtkeller65b}, to $112.3^\circ$ following \citet{Doering68} and $113^\circ$ following \citet{Elias11}. We analyze the origin of all these results and calculated the equilibrium value, about $105^\circ$, see \eqref{eq:Ep1}, which minimizes the total magnetostatic energy, not only the part of it.

The next problem appears in modeling of the Bloch point. As is was discussed by \citet{Thiaville03}, the modeling of singularity is mesh--dependent. In particular, a mesh--friction effect and a strong mesh dependence of the switching field during the Bloch--point--mediated vortex switching process was detected using \textsf{OOMMF} micromagnetic simulations.\cite{Thiaville03} The reason is that micromagnetic simulators consider the numerically discretized Landau--Lifstitz equation, which are valid in continuum theory. Since the Bloch point appears as a singularity of continuum theory, it is always located between mesh points, and causes the mesh--dependent effects and therefore may be insufficient for describing near--field Bloch point distribution. In contrast to this, spin--lattice simulations are free from these shortage. From the beginning we consider discrete spins, located on the cubic lattice, and their dynamics is governed by the discrete versions of Landau--Lifshitz equations. The lattice Hamiltonian allows us to calculate the discrete energy of the Bloch point similar to the atomiclike calculations by \citet{Reinhardt73}.

Using in--house developed spin--lattice \texttt{SLaSi}\cite{SLaSi} simulator we modeled the Bloch point state nanosphere and checked our analytical predictions about Bloch point structure. We stabilized the singularity inside the spherical particle by applied gradient magnetic field. The field causes the new type of Bloch point with radial--dependent rotation angle $\gamma(r)$.

\begin{acknowledgments}

Authors acknowledge computing time on the high--performance computing cluster of National Taras Shevchenko University of Kyiv \footnote{\url{http://cluster.univ.kiev.ua}} and SKIT-3 Computing Cluster of Glushkov Institute of Cybernetic of NAS of Ukraine \footnote{\url{http://icybcluster.org.ua/}}. This work was supported by the Grant of the President of Ukraine No. F35/538-2011. We thank V.~Kravchuk for helpful discussions.
\end{acknowledgments}

\appendix

\section{Bloch point structure in a weak field}
\label{sec:weak}

We consider here the magnetization structure of a Bloch point under the action of weak magnetic field. One has to linearize Eq.~\eqref{eq:varEqRad} on the background of the unperturbed rotation angle $\gamma_0$, see \eqref{eq:smallFieldsEstimation}, which can be presented as follows:
\begin{equation*} \label{eq:g}
\gamma(r) \approx \gamma_0 + b g_\varepsilon(r), \quad
g_\varepsilon(r) = \frac{2\sqrt{5\varepsilon}}{3} f(\lambda r),\quad \lambda = \frac{1}{2}\sqrt{\frac{3}{\varepsilon}}.
\end{equation*}
Here the function $f(\xi)$ satisfies the linearized version of Eq.~\eqref{eq:varEqRad}:
\begin{equation*} \label{eq:EqforSmallFields}
\frac{ \mathrm d^2f}{\mathrm d \xi^2} + \frac{2}{\xi} \frac{ \mathrm d f}{\mathrm d \xi}  - f = \xi,
\end{equation*}
which can be easily integrated:
\begin{equation} \label{eq:f(xi)}
\begin{split}
f(\xi)    &= C_\lambda \frac{\sinh \xi}{\xi} + 2\frac{\cosh\xi-1}{\xi } - \xi,\\
C_\lambda &= \frac{\lambda ^2-2 \lambda  \sinh \lambda +2 \cosh \lambda -2}{\lambda  \cosh\lambda -\sinh \lambda}.
\end{split}
\end{equation}
The graphics of the $g_\varepsilon(r)$ for $\varepsilon = 0.05$ is presented in Fig.~\ref{fig:fbyr-eps} together with numerical solution of Eq.~\eqref{eq:varEqRad} by shooting method. In spite of limitation of our analysis by the case of weak field, $|b|\ll1$, the function $g_\varepsilon(r)$ provides a good approximation for the solution of nonlinear Eq.~\eqref{eq:varEqRad} up to to very strong fields $b\le 1$ with a relative error $\bigl|\left[\gamma(r)_\text{num}-\gamma(r)_\text{theor}\right]/\gamma(r)_\text{num}\bigr| \le 0.04$. 
\begin{figure}
\includegraphics[width=\columnwidth]{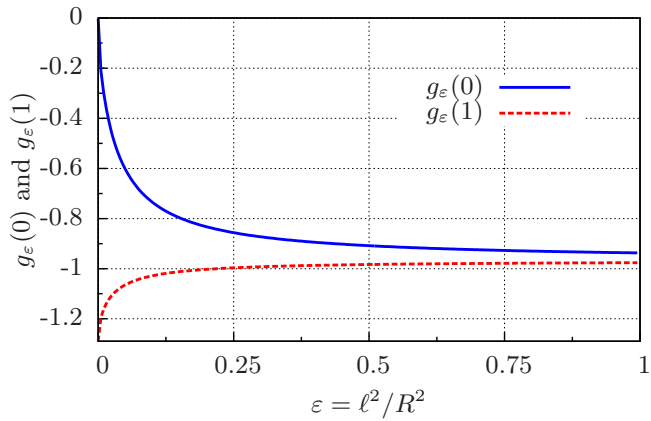}
\caption{(Color online.) Reduced rotation angle $g_\varepsilon$ \emph{vs} reduced exchange length $\varepsilon$: at $r=0$ (solid curve) and $r=1$ (dashed curve).}
\label{fig:inZeroAndOne}
\end{figure}

The rotation angle in the Bloch point is essentially influenced by the exchange parameter $\varepsilon$, see Fig.~\ref{fig:inZeroAndOne}. In the limit case of small particle ($\varepsilon\gg1$) the role of exchange is dominant, which  results in the constant angle $g_{\infty}=-\sqrt{15}/4\approx -0.97$. In the opposite case $\varepsilon \ll 1$, the role of magnetostatic interaction is enhanced and this leads to a nonhomogeneous rotational angle distribution. In the limiting case $g_0(0) = 0$ and $g_0(1) = -\sqrt{5/3}\approx-1.3$. Such a limit case is realized in typical soft nanomagnets sized in some tens of nanometers.

%

\begin{thebibliography}{45}%
\makeatletter
\providecommand \@ifxundefined [1]{%
 \@ifx{#1\undefined}
}%
\providecommand \@ifnum [1]{%
 \ifnum #1\expandafter \@firstoftwo
 \else \expandafter \@secondoftwo
 \fi
}%
\providecommand \@ifx [1]{%
 \ifx #1\expandafter \@firstoftwo
 \else \expandafter \@secondoftwo
 \fi
}%
\providecommand \natexlab [1]{#1}%
\providecommand \enquote  [1]{``#1''}%
\providecommand \bibnamefont  [1]{#1}%
\providecommand \bibfnamefont [1]{#1}%
\providecommand \citenamefont [1]{#1}%
\providecommand \href@noop [0]{\@secondoftwo}%
\providecommand \href [0]{\begingroup \@sanitize@url \@href}%
\providecommand \@href[1]{\@@startlink{#1}\@@href}%
\providecommand \@@href[1]{\endgroup#1\@@endlink}%
\providecommand \@sanitize@url [0]{\catcode `\\12\catcode `\$12\catcode
  `\&12\catcode `\#12\catcode `\^12\catcode `\_12\catcode `\%12\relax}%
\providecommand \@@startlink[1]{}%
\providecommand \@@endlink[0]{}%
\providecommand \url  [0]{\begingroup\@sanitize@url \@url }%
\providecommand \@url [1]{\endgroup\@href {#1}{\urlprefix }}%
\providecommand \urlprefix  [0]{URL }%
\providecommand \Eprint [0]{\href }%
\providecommand \doibase [0]{http://dx.doi.org/}%
\providecommand \selectlanguage [0]{\@gobble}%
\providecommand \bibinfo  [0]{\@secondoftwo}%
\providecommand \bibfield  [0]{\@secondoftwo}%
\providecommand \translation [1]{[#1]}%
\providecommand \BibitemOpen [0]{}%
\providecommand \bibitemStop [0]{}%
\providecommand \bibitemNoStop [0]{.\EOS\space}%
\providecommand \EOS [0]{\spacefactor3000\relax}%
\providecommand \BibitemShut  [1]{\csname bibitem#1\endcsname}%
\let\auto@bib@innerbib\@empty
\bibitem [{\citenamefont {Domb}\ and\ \citenamefont {Green}(1983)}]{Domb83}%
  \BibitemOpen
  \bibinfo {editor} {\bibfnamefont {C.}~\bibnamefont {Domb}}\ and\ \bibinfo
  {editor} {\bibfnamefont {M.~S.}\ \bibnamefont {Green}},\ eds.,\ \href@noop {}
  {\emph {\bibinfo {title} {Phase Transitions and Critical Phenomena}}},\
  Vol.~\bibinfo {volume} {7}\ (\bibinfo  {publisher} {Academic Press},\
  \bibinfo {year} {1983})\ p.\ \bibinfo {pages} {328}\BibitemShut {NoStop}%
\bibitem [{\citenamefont {Strandburg}(1991)}]{Strandburg91}%
  \BibitemOpen
  \bibinfo {editor} {\bibfnamefont {K.~J.}\ \bibnamefont {Strandburg}},\ ed.,\
  \href@noop {} {\emph {\bibinfo {title} {Bond--Orientational Order in
  Condensed Matter Systems}}}\ (\bibinfo  {publisher} {Springer},\ \bibinfo
  {year} {1991})\ p.\ \bibinfo {pages} {388}\BibitemShut {NoStop}%
\bibitem [{\citenamefont {Yonezawa}\ and\ \citenamefont
  {Ninomiya}(1983)}]{Yonezawa83}%
  \BibitemOpen
  \bibfield  {author} {\bibinfo {author} {\bibfnamefont {F.}~\bibnamefont
  {Yonezawa}}\ and\ \bibinfo {author} {\bibfnamefont {T.}~\bibnamefont
  {Ninomiya}},\ }\href@noop {} {\emph {\bibinfo {title} {Topological Disorder
  in Condensed Matter}}},\ \bibinfo {series} {Springer series in solid-state
  sciences}, Vol.~\bibinfo {volume} {46}\ (\bibinfo  {publisher} {Springer},\
  \bibinfo {year} {1983})\ p.\ \bibinfo {pages} {253}\BibitemShut {NoStop}%
\bibitem [{\citenamefont {Bowick}\ and\ \citenamefont
  {Giomi}(2009)}]{Bowick09}%
  \BibitemOpen
  \bibfield  {author} {\bibinfo {author} {\bibfnamefont {M.~J.}\ \bibnamefont
  {Bowick}}\ and\ \bibinfo {author} {\bibfnamefont {L.}~\bibnamefont {Giomi}},\
  }\bibfield  {title} {\enquote {\bibinfo {title} {Two-dimensional matter:
  order, curvature and defects},}\ }\href {\doibase 10.1080/00018730903043166}
  {\bibfield  {journal} {\bibinfo  {journal} {Advances in Physics}\ }\textbf
  {\bibinfo {volume} {58}},\ \bibinfo {pages} {449--563} (\bibinfo {year}
  {2009})}\BibitemShut {NoStop}%
\bibitem [{\citenamefont {Fetter}(2009)}]{Fetter09}%
  \BibitemOpen
  \bibfield  {author} {\bibinfo {author} {\bibfnamefont {A.~L.}\ \bibnamefont
  {Fetter}},\ }\bibfield  {title} {\enquote {\bibinfo {title} {Rotating trapped
  bose-einstein condensates},}\ }\href {\doibase 10.1103/RevModPhys.81.647}
  {\bibfield  {journal} {\bibinfo  {journal} {Rev. Mod. Phys.}\ }\textbf
  {\bibinfo {volume} {81}},\ \bibinfo {pages} {647--691} (\bibinfo {year}
  {2009})}\BibitemShut {NoStop}%
\bibitem [{\citenamefont {Hubert}\ and\ \citenamefont {Sch{\"
  a}fer}(1998)}]{Hubert98}%
  \BibitemOpen
  \bibfield  {author} {\bibinfo {author} {\bibfnamefont {A.}~\bibnamefont
  {Hubert}}\ and\ \bibinfo {author} {\bibfnamefont {R.}~\bibnamefont {Sch{\"
  a}fer}},\ }\href@noop {} {\emph {\bibinfo {title} {Magnetic domains: the
  analysis of magnetic microstructures}}}\ (\bibinfo  {publisher}
  {Springer--Verlag},\ \bibinfo {address} {Berlin},\ \bibinfo {year}
  {1998})\BibitemShut {NoStop}%
\bibitem [{\citenamefont {St{\"o}hr}\ and\ \citenamefont
  {Siegmann}(2006)}]{Stohr06}%
  \BibitemOpen
  \bibfield  {author} {\bibinfo {author} {\bibfnamefont {J.}~\bibnamefont
  {St{\"o}hr}}\ and\ \bibinfo {author} {\bibfnamefont {H.~C.}\ \bibnamefont
  {Siegmann}},\ }\href@noop {} {\emph {\bibinfo {title} {Magnetism: From
  Fundamentals to Nanoscale Dynamics}}},\ \bibinfo {series} {Springer Series in
  solid-state sciences}, Vol.\ \bibinfo {volume} {152}\ (\bibinfo  {publisher}
  {Springer-Verlag Berlin Heidelberg},\ \bibinfo {year} {2006})\BibitemShut
  {NoStop}%
\bibitem [{\citenamefont {Guimar$\tilde{\mathrm{a}}$es}(2009)}]{Guimaraes09}%
  \BibitemOpen
  \bibfield  {author} {\bibinfo {author} {\bibfnamefont {A.~P.}\ \bibnamefont
  {Guimar$\tilde{\mathrm{a}}$es}},\ }\href {\doibase 10.1007/978-3-642-01482-6}
  {\emph {\bibinfo {title} {Principles of Nanomagnetism}}},\ NanoScience and
  Technology\ (\bibinfo  {publisher} {Springer-Verlag Berlin Heidelberg},\
  \bibinfo {year} {2009})\BibitemShut {NoStop}%
\bibitem [{\citenamefont {Naumov}, \citenamefont {Bellaiche},\ and\
  \citenamefont {Fu}(2004)}]{Naumov04}%
  \BibitemOpen
  \bibfield  {author} {\bibinfo {author} {\bibfnamefont {I.~I.}\ \bibnamefont
  {Naumov}}, \bibinfo {author} {\bibfnamefont {L.}~\bibnamefont {Bellaiche}}, \
  and\ \bibinfo {author} {\bibfnamefont {H.}~\bibnamefont {Fu}},\ }\bibfield
  {title} {\enquote {\bibinfo {title} {Unusual phase transitions in
  ferroelectric nanodisks and nanorods},}\ }\href
  {http://dx.doi.org/10.1038/nature03107} {\bibfield  {journal} {\bibinfo
  {journal} {Nature}\ }\textbf {\bibinfo {volume} {432}},\ \bibinfo {pages}
  {737--740} (\bibinfo {year} {2004})}\BibitemShut {NoStop}%
\bibitem [{\citenamefont {Senthil}\ \emph {et~al.}(2004)\citenamefont
  {Senthil}, \citenamefont {Vishwanath}, \citenamefont {Balents}, \citenamefont
  {Sachdev},\ and\ \citenamefont {Fisher}}]{Senthil04}%
  \BibitemOpen
  \bibfield  {author} {\bibinfo {author} {\bibfnamefont {T.}~\bibnamefont
  {Senthil}}, \bibinfo {author} {\bibfnamefont {A.}~\bibnamefont {Vishwanath}},
  \bibinfo {author} {\bibfnamefont {L.}~\bibnamefont {Balents}}, \bibinfo
  {author} {\bibfnamefont {S.}~\bibnamefont {Sachdev}}, \ and\ \bibinfo
  {author} {\bibfnamefont {M.~P.~A.}\ \bibnamefont {Fisher}},\ }\bibfield
  {title} {\enquote {\bibinfo {title} {Deconfined quantum critical points},}\
  }\href {\doibase 10.1126/science.1091806} {\bibfield  {journal} {\bibinfo
  {journal} {Science}\ }\textbf {\bibinfo {volume} {303}},\ \bibinfo {pages}
  {1490--1494} (\bibinfo {year} {2004})},\ \Eprint
  {http://arxiv.org/abs/http://www.sciencemag.org/content/303/5663/1490.full.pdf}
  {http://www.sciencemag.org/content/303/5663/1490.full.pdf} \BibitemShut
  {NoStop}%
\bibitem [{\citenamefont {Mermin}(1981)}]{Mermin81}%
  \BibitemOpen
  \bibfield  {author} {\bibinfo {author} {\bibnamefont {Mermin}},\ }\bibfield
  {title} {\enquote {\bibinfo {title} {E pluribus boojum: the physicist as
  neologist},}\ }\href@noop {} {\bibfield  {journal} {\bibinfo  {journal}
  {Physics Today}\ }\textbf {\bibinfo {volume} {34}},\ \bibinfo {pages} {46–53}
  (\bibinfo {year} {1981})}\BibitemShut {NoStop}%
\bibitem [{\citenamefont {Volovik}(2003)}]{Volovik03}%
  \BibitemOpen
  \bibfield  {author} {\bibinfo {author} {\bibfnamefont {G.}~\bibnamefont
  {Volovik}},\ }\href
  {http://www.zentralblatt-math.org/zmath/search/?an=01866310} {\emph {\bibinfo
  {title} {The universe in a {H}elium droplet}}}\ (\bibinfo  {publisher}
  {Oxford University Press},\ \bibinfo {address} {Oxford},\ \bibinfo {year}
  {2003})\BibitemShut {NoStop}%
\bibitem [{\citenamefont {Malozemoff}\ and\ \citenamefont
  {Slonzewski}(1979)}]{Malozemoff79}%
  \BibitemOpen
  \bibfield  {author} {\bibinfo {author} {\bibfnamefont {A.~P.}\ \bibnamefont
  {Malozemoff}}\ and\ \bibinfo {author} {\bibfnamefont {J.~C.}\ \bibnamefont
  {Slonzewski}},\ }\href@noop {} {\emph {\bibinfo {title} {Magnetic domain
  walls in bubble materials}}}\ (\bibinfo  {publisher} {Academic Press},\
  \bibinfo {address} {New York},\ \bibinfo {year} {1979})\BibitemShut {NoStop}%
\bibitem [{\citenamefont {Feldtkeller}(1965)}]{Feldtkeller65b}%
  \BibitemOpen
  \bibfield  {author} {\bibinfo {author} {\bibfnamefont {E.}~\bibnamefont
  {Feldtkeller}},\ }\bibfield  {title} {\enquote {\bibinfo {title}
  {Mikromagnetisch stetige und unstetige magnetisierungskonfigurationen},}\
  }\href@noop {} {\bibfield  {journal} {\bibinfo  {journal} {Zeitschrift
  f\"{u}r angewandte Physik}\ }\textbf {\bibinfo {volume} {19}},\ \bibinfo
  {pages} {530--536} (\bibinfo {year} {1965})}\BibitemShut {NoStop}%
\bibitem [{\citenamefont {D\"{o}ring}(1968)}]{Doering68}%
  \BibitemOpen
  \bibfield  {author} {\bibinfo {author} {\bibfnamefont {W.}~\bibnamefont
  {D\"{o}ring}},\ }\bibfield  {title} {\enquote {\bibinfo {title} {Point
  singularities in micromagnetism},}\ }\href {\doibase 10.1063/1.1656144}
  {\bibfield  {journal} {\bibinfo  {journal} {J.~Appl. Phys.}\ }\textbf
  {\bibinfo {volume} {39}},\ \bibinfo {pages} {1006--1007} (\bibinfo {year}
  {1968})}\BibitemShut {NoStop}%
\bibitem [{\citenamefont {Kabanov}, \citenamefont {Dedukh},\ and\ \citenamefont
  {Nikitenko}(1989)}]{Kabanov89}%
  \BibitemOpen
  \bibfield  {author} {\bibinfo {author} {\bibfnamefont {Y.~P.}\ \bibnamefont
  {Kabanov}}, \bibinfo {author} {\bibfnamefont {L.~M.}\ \bibnamefont {Dedukh}},
  \ and\ \bibinfo {author} {\bibfnamefont {V.~I.}\ \bibnamefont {Nikitenko}},\
  }\bibfield  {title} {\enquote {\bibinfo {title} {Bloch points in an
  oscillating bloch line},}\ }\href
  {http://www.jetpletters.ac.ru/ps/1121/article_16982.shtml} {\bibfield
  {journal} {\bibinfo  {journal} {JETP Lett}\ }\textbf {\bibinfo {volume}
  {49}},\ \bibinfo {pages} {637} (\bibinfo {year} {1989})}\BibitemShut
  {NoStop}%
\bibitem [{\citenamefont {Hertel}\ and\ \citenamefont
  {Kirschner}(2004)}]{Hertel04a}%
  \BibitemOpen
  \bibfield  {author} {\bibinfo {author} {\bibfnamefont {R.}~\bibnamefont
  {Hertel}}\ and\ \bibinfo {author} {\bibfnamefont {J.}~\bibnamefont
  {Kirschner}},\ }\bibfield  {title} {\enquote {\bibinfo {title} {Magnetic
  drops in a soft-magnetic cylinder},}\ }\href
  {http://www.sciencedirect.com/science/article/B6TJJ-4BYP1W4-4/2/f85731b704a46f19c401b90833e467b7}
  {\bibfield  {journal} {\bibinfo  {journal} {J.~Magn. Magn. Mater.}\ }\textbf
  {\bibinfo {volume} {278}},\ \bibinfo {pages} {L291--L297} (\bibinfo {year}
  {2004})}\BibitemShut {NoStop}%
\bibitem [{\citenamefont {Porrati}\ and\ \citenamefont
  {Huth}(2005)}]{Porrati05}%
  \BibitemOpen
  \bibfield  {author} {\bibinfo {author} {\bibfnamefont {F.}~\bibnamefont
  {Porrati}}\ and\ \bibinfo {author} {\bibfnamefont {M.}~\bibnamefont {Huth}},\
  }\bibfield  {title} {\enquote {\bibinfo {title} {Micromagnetic structure and
  vortex core reversal in arrays of iron nano-cylinders},}\ }\bibfield
  {booktitle} {\emph {\bibinfo {booktitle} {Proceedings of the Joint European
  Magnetic Symposia (JEMS' 04)}},\ }\href
  {http://www.sciencedirect.com/science/article/B6TJJ-4DXT3YT-5/2/b071400d1def405a08ece5113cf1f28f}
  {\bibfield  {journal} {\bibinfo  {journal} {J.~Magn. Magn. Mater.}\ }\textbf
  {\bibinfo {volume} {290-291}},\ \bibinfo {pages} {145--148} (\bibinfo {year}
  {2005})}\BibitemShut {NoStop}%
\bibitem [{\citenamefont {Niedoba}\ and\ \citenamefont
  {Labrune}(2005)}]{Niedoba05}%
  \BibitemOpen
  \bibfield  {author} {\bibinfo {author} {\bibfnamefont {H.}~\bibnamefont
  {Niedoba}}\ and\ \bibinfo {author} {\bibfnamefont {M.}~\bibnamefont
  {Labrune}},\ }\bibfield  {title} {\enquote {\bibinfo {title} {Magnetization
  reversal via bloch points nucleation in nanowires and dots: a micromagnetic
  study},}\ }\href {http://dx.doi.org/10.1140/epjb/e2005-00353-6} {\bibfield
  {journal} {\bibinfo  {journal} {The European Physical Journal B - Condensed
  Matter and Complex Systems}\ }\textbf {\bibinfo {volume} {V47}},\ \bibinfo
  {pages} {467--478} (\bibinfo {year} {2005})}\BibitemShut {NoStop}%
\bibitem [{\citenamefont {Vila}\ \emph {et~al.}(2009)\citenamefont {Vila},
  \citenamefont {Darques}, \citenamefont {Encinas}, \citenamefont {Ebels},
  \citenamefont {George}, \citenamefont {Faini}, \citenamefont {Thiaville},\
  and\ \citenamefont {Piraux}}]{Vila09}%
  \BibitemOpen
  \bibfield  {author} {\bibinfo {author} {\bibfnamefont {L.}~\bibnamefont
  {Vila}}, \bibinfo {author} {\bibfnamefont {M.}~\bibnamefont {Darques}},
  \bibinfo {author} {\bibfnamefont {A.}~\bibnamefont {Encinas}}, \bibinfo
  {author} {\bibfnamefont {U.}~\bibnamefont {Ebels}}, \bibinfo {author}
  {\bibfnamefont {J.-M.}\ \bibnamefont {George}}, \bibinfo {author}
  {\bibfnamefont {G.}~\bibnamefont {Faini}}, \bibinfo {author} {\bibfnamefont
  {A.}~\bibnamefont {Thiaville}}, \ and\ \bibinfo {author} {\bibfnamefont
  {L.}~\bibnamefont {Piraux}},\ }\bibfield  {title} {\enquote {\bibinfo {title}
  {Magnetic vortices in nanowires with transverse easy axis},}\ }\href
  {\doibase 10.1103/PhysRevB.79.172410} {\bibfield  {journal} {\bibinfo
  {journal} {Phys. Rev. B}\ }\textbf {\bibinfo {volume} {79}},\ \bibinfo {eid}
  {172410} (\bibinfo {year} {2009})}\BibitemShut {NoStop}%
\bibitem [{\citenamefont {Masseboeuf}\ \emph {et~al.}(2009)\citenamefont
  {Masseboeuf}, \citenamefont {Jourdan}, \citenamefont {Lancon}, \citenamefont
  {Bayle-Guillemaud},\ and\ \citenamefont {Marty}}]{Masseboeuf09}%
  \BibitemOpen
  \bibfield  {author} {\bibinfo {author} {\bibfnamefont {A.}~\bibnamefont
  {Masseboeuf}}, \bibinfo {author} {\bibfnamefont {T.}~\bibnamefont {Jourdan}},
  \bibinfo {author} {\bibfnamefont {F.}~\bibnamefont {Lancon}}, \bibinfo
  {author} {\bibfnamefont {P.}~\bibnamefont {Bayle-Guillemaud}}, \ and\
  \bibinfo {author} {\bibfnamefont {A.}~\bibnamefont {Marty}},\ }\bibfield
  {title} {\enquote {\bibinfo {title} {Probing magnetic singularities during
  magnetization process in fepd films},}\ }\href {\doibase 10.1063/1.3266825}
  {\bibfield  {journal} {\bibinfo  {journal} {Appl. Phys. Lett.}\ }\textbf
  {\bibinfo {volume} {95}},\ \bibinfo {eid} {212501} (\bibinfo {year}
  {2009})}\BibitemShut {NoStop}%
\bibitem [{\citenamefont {Jourdan}\ \emph {et~al.}(2009)\citenamefont
  {Jourdan}, \citenamefont {Masseboeuf}, \citenamefont {Lan\c{c}on},
  \citenamefont {Bayle-Guillemaud},\ and\ \citenamefont {Marty}}]{Jourdan09}%
  \BibitemOpen
  \bibfield  {author} {\bibinfo {author} {\bibfnamefont {T.}~\bibnamefont
  {Jourdan}}, \bibinfo {author} {\bibfnamefont {A.}~\bibnamefont {Masseboeuf}},
  \bibinfo {author} {\bibfnamefont {F.}~\bibnamefont {Lan\c{c}on}}, \bibinfo
  {author} {\bibfnamefont {P.}~\bibnamefont {Bayle-Guillemaud}}, \ and\
  \bibinfo {author} {\bibfnamefont {A.}~\bibnamefont {Marty}},\ }\bibfield
  {title} {\enquote {\bibinfo {title} {Magnetic bubbles in fepd thin films near
  saturation},}\ }\href {\doibase 10.1063/1.3243318} {\bibfield  {journal}
  {\bibinfo  {journal} {J.~Appl. Phys.}\ }\textbf {\bibinfo {volume} {106}},\
  \bibinfo {eid} {073913} (\bibinfo {year} {2009})}\BibitemShut {NoStop}%
\bibitem [{\citenamefont {Thiaville}\ \emph {et~al.}(2003)\citenamefont
  {Thiaville}, \citenamefont {Garcia}, \citenamefont {Dittrich}, \citenamefont
  {Miltat},\ and\ \citenamefont {Schrefl}}]{Thiaville03}%
  \BibitemOpen
  \bibfield  {author} {\bibinfo {author} {\bibfnamefont {A.}~\bibnamefont
  {Thiaville}}, \bibinfo {author} {\bibfnamefont {J.~M.}\ \bibnamefont
  {Garcia}}, \bibinfo {author} {\bibfnamefont {R.}~\bibnamefont {Dittrich}},
  \bibinfo {author} {\bibfnamefont {J.}~\bibnamefont {Miltat}}, \ and\ \bibinfo
  {author} {\bibfnamefont {T.}~\bibnamefont {Schrefl}},\ }\bibfield  {title}
  {\enquote {\bibinfo {title} {Micromagnetic study of bloch-point-mediated
  vortex core reversal},}\ }\href
  {http://link.aps.org/abstract/PRB/v67/e094410} {\bibfield  {journal}
  {\bibinfo  {journal} {Phys. Rev. B}\ }\textbf {\bibinfo {volume} {67}},\
  \bibinfo {eid} {094410} (\bibinfo {year} {2003})}\BibitemShut {NoStop}%
\bibitem [{\citenamefont {Hertel}\ and\ \citenamefont
  {Schneider}(2006)}]{Hertel06}%
  \BibitemOpen
  \bibfield  {author} {\bibinfo {author} {\bibfnamefont {R.}~\bibnamefont
  {Hertel}}\ and\ \bibinfo {author} {\bibfnamefont {C.~M.}\ \bibnamefont
  {Schneider}},\ }\bibfield  {title} {\enquote {\bibinfo {title} {Exchange
  explosions: Magnetization dynamics during vortex-antivortex annihilation},}\
  }\href {http://link.aps.org/abstract/PRL/v97/e177202} {\bibfield  {journal}
  {\bibinfo  {journal} {Phys. Rev. Lett.}\ }\textbf {\bibinfo {volume} {97}},\
  \bibinfo {eid} {177202} (\bibinfo {year} {2006})}\BibitemShut {NoStop}%
\bibitem [{\citenamefont {Xing}\ \emph {et~al.}(2008)\citenamefont {Xing},
  \citenamefont {Yu}, \citenamefont {Wu}, \citenamefont {Xu},\ and\
  \citenamefont {Li}}]{Xing08}%
  \BibitemOpen
  \bibfield  {author} {\bibinfo {author} {\bibfnamefont {X.~J.}\ \bibnamefont
  {Xing}}, \bibinfo {author} {\bibfnamefont {Y.~P.}\ \bibnamefont {Yu}},
  \bibinfo {author} {\bibfnamefont {S.~X.}\ \bibnamefont {Wu}}, \bibinfo
  {author} {\bibfnamefont {L.~M.}\ \bibnamefont {Xu}}, \ and\ \bibinfo {author}
  {\bibfnamefont {S.~W.}\ \bibnamefont {Li}},\ }\bibfield  {title} {\enquote
  {\bibinfo {title} {Bloch-point-mediated magnetic antivortex core reversal
  triggered by sudden excitation of a suprathreshold spin-polarized current},}\
  }\href {\doibase 10.1063/1.3033400} {\bibfield  {journal} {\bibinfo
  {journal} {Appl. Phys. Lett.}\ }\textbf {\bibinfo {volume} {93}},\ \bibinfo
  {eid} {202507} (\bibinfo {year} {2008})}\BibitemShut {NoStop}%
\bibitem [{\citenamefont {Okuno}\ \emph {et~al.}(2002)\citenamefont {Okuno},
  \citenamefont {Shigeto}, \citenamefont {Ono}, \citenamefont {Mibu},\ and\
  \citenamefont {Shinjo}}]{Okuno02}%
  \BibitemOpen
  \bibfield  {author} {\bibinfo {author} {\bibfnamefont {T.}~\bibnamefont
  {Okuno}}, \bibinfo {author} {\bibfnamefont {K.}~\bibnamefont {Shigeto}},
  \bibinfo {author} {\bibfnamefont {T.}~\bibnamefont {Ono}}, \bibinfo {author}
  {\bibfnamefont {K.}~\bibnamefont {Mibu}}, \ and\ \bibinfo {author}
  {\bibfnamefont {T.}~\bibnamefont {Shinjo}},\ }\bibfield  {title} {\enquote
  {\bibinfo {title} {{MFM} study of magnetic vortex cores in circular permalloy
  dots: behavior in external field},}\ }\href
  {http://www.sciencedirect.com/science/article/B6TJJ-447DCMV-3/2/84e9fdfbf9e0cab3d3182fc4db4b4032}
  {\bibfield  {journal} {\bibinfo  {journal} {J.~Magn. Magn. Mater.}\ }\textbf
  {\bibinfo {volume} {240}},\ \bibinfo {pages} {1--6} (\bibinfo {year}
  {2002})}\BibitemShut {NoStop}%
\bibitem [{\citenamefont {Kravchuk}\ and\ \citenamefont
  {Sheka}(2007)}]{Kravchuk07a}%
  \BibitemOpen
  \bibfield  {author} {\bibinfo {author} {\bibfnamefont {V.}~\bibnamefont
  {Kravchuk}}\ and\ \bibinfo {author} {\bibfnamefont {D.}~\bibnamefont
  {Sheka}},\ }\bibfield  {title} {\enquote {\bibinfo {title} {Thin
  ferromagnetic nanodisk in transverse magnetic field},}\ }\href
  {http://dx.doi.org/10.1134/S1063783407100186} {\bibfield  {journal} {\bibinfo
   {journal} {Physics of the Solid State}\ }\textbf {\bibinfo {volume} {49}},\
  \bibinfo {pages} {1923--1931} (\bibinfo {year} {2007})}\BibitemShut {NoStop}%
\bibitem [{\citenamefont {Van~Waeyenberge}\ \emph {et~al.}(2006)\citenamefont
  {Van~Waeyenberge}, \citenamefont {Puzic}, \citenamefont {Stoll},
  \citenamefont {Chou}, \citenamefont {Tyliszczak}, \citenamefont {Hertel},
  \citenamefont {F\"ahnle}, \citenamefont {Bruckl}, \citenamefont {Rott},
  \citenamefont {Reiss}, \citenamefont {Neudecker}, \citenamefont {Weiss},
  \citenamefont {Back},\ and\ \citenamefont {Sch\"utz}}]{Waeyenberge06}%
  \BibitemOpen
  \bibfield  {author} {\bibinfo {author} {\bibfnamefont {B.}~\bibnamefont
  {Van~Waeyenberge}}, \bibinfo {author} {\bibfnamefont {A.}~\bibnamefont
  {Puzic}}, \bibinfo {author} {\bibfnamefont {H.}~\bibnamefont {Stoll}},
  \bibinfo {author} {\bibfnamefont {K.~W.}\ \bibnamefont {Chou}}, \bibinfo
  {author} {\bibfnamefont {T.}~\bibnamefont {Tyliszczak}}, \bibinfo {author}
  {\bibfnamefont {R.}~\bibnamefont {Hertel}}, \bibinfo {author} {\bibfnamefont
  {M.}~\bibnamefont {F\"ahnle}}, \bibinfo {author} {\bibfnamefont
  {H.}~\bibnamefont {Bruckl}}, \bibinfo {author} {\bibfnamefont
  {K.}~\bibnamefont {Rott}}, \bibinfo {author} {\bibfnamefont {G.}~\bibnamefont
  {Reiss}}, \bibinfo {author} {\bibfnamefont {I.}~\bibnamefont {Neudecker}},
  \bibinfo {author} {\bibfnamefont {D.}~\bibnamefont {Weiss}}, \bibinfo
  {author} {\bibfnamefont {C.~H.}\ \bibnamefont {Back}}, \ and\ \bibinfo
  {author} {\bibfnamefont {G.}~\bibnamefont {Sch\"utz}},\ }\bibfield  {title}
  {\enquote {\bibinfo {title} {Magnetic vortex core reversal by excitation with
  short bursts of an alternating field},}\ }\href
  {http://dx.doi.org/10.1038/nature05240} {\bibfield  {journal} {\bibinfo
  {journal} {Nature}\ }\textbf {\bibinfo {volume} {444}},\ \bibinfo {pages}
  {461--464} (\bibinfo {year} {2006})}\BibitemShut {NoStop}%
\bibitem [{\citenamefont {Xiao}\ \emph {et~al.}(2006)\citenamefont {Xiao},
  \citenamefont {Rudge}, \citenamefont {Choi}, \citenamefont {Hong},\ and\
  \citenamefont {Donohoe}}]{Xiao06}%
  \BibitemOpen
  \bibfield  {author} {\bibinfo {author} {\bibfnamefont {Q.~F.}\ \bibnamefont
  {Xiao}}, \bibinfo {author} {\bibfnamefont {J.}~\bibnamefont {Rudge}},
  \bibinfo {author} {\bibfnamefont {B.~C.}\ \bibnamefont {Choi}}, \bibinfo
  {author} {\bibfnamefont {Y.~K.}\ \bibnamefont {Hong}}, \ and\ \bibinfo
  {author} {\bibfnamefont {G.}~\bibnamefont {Donohoe}},\ }\bibfield  {title}
  {\enquote {\bibinfo {title} {Dynamics of vortex core switching in
  ferromagnetic nanodisks},}\ }\href
  {http://link.aip.org/link/?APL/89/262507/1} {\bibfield  {journal} {\bibinfo
  {journal} {Appl. Phys. Lett.}\ }\textbf {\bibinfo {volume} {89}},\ \bibinfo
  {eid} {262507} (\bibinfo {year} {2006})}\BibitemShut {NoStop}%
\bibitem [{\citenamefont {Hertel}\ \emph {et~al.}(2007)\citenamefont {Hertel},
  \citenamefont {Gliga}, \citenamefont {F\"ahnle},\ and\ \citenamefont
  {Schneider}}]{Hertel07}%
  \BibitemOpen
  \bibfield  {author} {\bibinfo {author} {\bibfnamefont {R.}~\bibnamefont
  {Hertel}}, \bibinfo {author} {\bibfnamefont {S.}~\bibnamefont {Gliga}},
  \bibinfo {author} {\bibfnamefont {M.}~\bibnamefont {F\"ahnle}}, \ and\
  \bibinfo {author} {\bibfnamefont {C.~M.}\ \bibnamefont {Schneider}},\
  }\bibfield  {title} {\enquote {\bibinfo {title} {Ultrafast nanomagnetic
  toggle switching of vortex cores},}\ }\href
  {http://link.aps.org/abstract/PRL/v98/e117201} {\bibfield  {journal}
  {\bibinfo  {journal} {Phys. Rev. Lett.}\ }\textbf {\bibinfo {volume} {98}},\
  \bibinfo {eid} {117201} (\bibinfo {year} {2007})}\BibitemShut {NoStop}%
\bibitem [{\citenamefont {Yamada}\ \emph {et~al.}(2007)\citenamefont {Yamada},
  \citenamefont {Kasai}, \citenamefont {Nakatani}, \citenamefont {Kobayashi},
  \citenamefont {Kohno}, \citenamefont {Thiaville},\ and\ \citenamefont
  {Ono}}]{Yamada07}%
  \BibitemOpen
  \bibfield  {author} {\bibinfo {author} {\bibfnamefont {K.}~\bibnamefont
  {Yamada}}, \bibinfo {author} {\bibfnamefont {S.}~\bibnamefont {Kasai}},
  \bibinfo {author} {\bibfnamefont {Y.}~\bibnamefont {Nakatani}}, \bibinfo
  {author} {\bibfnamefont {K.}~\bibnamefont {Kobayashi}}, \bibinfo {author}
  {\bibfnamefont {H.}~\bibnamefont {Kohno}}, \bibinfo {author} {\bibfnamefont
  {A.}~\bibnamefont {Thiaville}}, \ and\ \bibinfo {author} {\bibfnamefont
  {T.}~\bibnamefont {Ono}},\ }\bibfield  {title} {\enquote {\bibinfo {title}
  {Electrical switching of the vortex core in a magnetic disk},}\ }\href
  {http://dx.doi.org/10.1038/nmat1867} {\bibfield  {journal} {\bibinfo
  {journal} {Nat Mater}\ }\textbf {\bibinfo {volume} {6}},\ \bibinfo {pages}
  {270--273} (\bibinfo {year} {2007})}\BibitemShut {NoStop}%
\bibitem [{\citenamefont {Kravchuk}\ \emph {et~al.}(2007)\citenamefont
  {Kravchuk}, \citenamefont {Sheka}, \citenamefont {Gaididei},\ and\
  \citenamefont {Mertens}}]{Kravchuk07c}%
  \BibitemOpen
  \bibfield  {author} {\bibinfo {author} {\bibfnamefont {V.~P.}\ \bibnamefont
  {Kravchuk}}, \bibinfo {author} {\bibfnamefont {D.~D.}\ \bibnamefont {Sheka}},
  \bibinfo {author} {\bibfnamefont {Y.}~\bibnamefont {Gaididei}}, \ and\
  \bibinfo {author} {\bibfnamefont {F.~G.}\ \bibnamefont {Mertens}},\
  }\bibfield  {title} {\enquote {\bibinfo {title} {Controlled vortex core
  switching in a magnetic nanodisk by a rotating field},}\ }\href {\doibase
  10.1063/1.2770819} {\bibfield  {journal} {\bibinfo  {journal} {J.~Appl.
  Phys.}\ }\textbf {\bibinfo {volume} {102}},\ \bibinfo {eid} {043908}
  (\bibinfo {year} {2007})}\BibitemShut {NoStop}%
\bibitem [{\citenamefont {Sheka}, \citenamefont {Gaididei},\ and\ \citenamefont
  {Mertens}(2007)}]{Sheka07b}%
  \BibitemOpen
  \bibfield  {author} {\bibinfo {author} {\bibfnamefont {D.~D.}\ \bibnamefont
  {Sheka}}, \bibinfo {author} {\bibfnamefont {Y.}~\bibnamefont {Gaididei}}, \
  and\ \bibinfo {author} {\bibfnamefont {F.~G.}\ \bibnamefont {Mertens}},\
  }\bibfield  {title} {\enquote {\bibinfo {title} {Current induced switching of
  vortex polarity in magnetic nanodisks},}\ }\href {\doibase 10.1063/1.2775036}
  {\bibfield  {journal} {\bibinfo  {journal} {Appl. Phys. Lett.}\ }\textbf
  {\bibinfo {volume} {91}},\ \bibinfo {pages} {082509} (\bibinfo {year}
  {2007})}\BibitemShut {NoStop}%
\bibitem [{\citenamefont {Galkina}\ and\ \citenamefont
  {Ivanov}(1995)}]{Galkina95}%
  \BibitemOpen
  \bibfield  {author} {\bibinfo {author} {\bibfnamefont {E.~G.}\ \bibnamefont
  {Galkina}}\ and\ \bibinfo {author} {\bibfnamefont {B.~A.}\ \bibnamefont
  {Ivanov}},\ }\bibfield  {title} {\enquote {\bibinfo {title} {Quantum
  tunneling in a magnetic vortex in a 2d easy-plane magnetic material},}\
  }\href {http://www.jetpletters.ac.ru/ps/1205/article_18218.shtml} {\bibfield
  {journal} {\bibinfo  {journal} {JETP Lett.}\ }\textbf {\bibinfo {volume}
  {61}},\ \bibinfo {pages} {511--514} (\bibinfo {year} {1995})}\BibitemShut
  {NoStop}%
\bibitem [{Note1()}]{Note1}%
  \BibitemOpen
  \bibinfo {note} {Bloch points in magnetic bubbles were classified by \protect
  \citet {Malozemoff79} using the flux $N$ of gyrotropic vector. Simple
  calculations show that the value of such a flux is opposite to the
  topological density \protect \textup {\hbox {\mathsurround \z@ \protect
  \normalfont (\ignorespaces \ref {eq:Pontryagin}\unskip \@@italiccorr )}}, $N
  =-Q$.}\BibitemShut {Stop}%
\bibitem [{\citenamefont {Aharoni}(1996)}]{Aharoni96}%
  \BibitemOpen
  \bibfield  {author} {\bibinfo {author} {\bibfnamefont {A.}~\bibnamefont
  {Aharoni}},\ }\href@noop {} {\emph {\bibinfo {title} {Introduction to the
  theory of {F}erromagnetism}}}\ (\bibinfo  {publisher} {Oxford University
  Press},\ \bibinfo {year} {1996})\BibitemShut {NoStop}%
\bibitem [{\citenamefont {El{\'i}as}\ and\ \citenamefont
  {Verga}(2011)}]{Elias11}%
  \BibitemOpen
  \bibfield  {author} {\bibinfo {author} {\bibfnamefont {R.}~\bibnamefont
  {El{\'i}as}}\ and\ \bibinfo {author} {\bibfnamefont {A.}~\bibnamefont
  {Verga}},\ }\bibfield  {title} {\enquote {\bibinfo {title} {Magnetization
  structure of a bloch point singularity},}\ }\href
  {http://dx.doi.org/10.1140/epjb/e2011-20146-6} {\bibfield  {journal}
  {\bibinfo  {journal} {The European Physical Journal B - Condensed Matter and
  Complex Systems}\ ,\ \bibinfo {pages} {1--8}} (\bibinfo {year} {2011})},\
  \bibinfo {note} {10.1140/epjb/e2011-20146-6}\BibitemShut {NoStop}%
\bibitem [{\citenamefont {Manton}\ and\ \citenamefont
  {Sutcliffe}(2004)}]{Manton04}%
  \BibitemOpen
  \bibfield  {author} {\bibinfo {author} {\bibfnamefont {N.}~\bibnamefont
  {Manton}}\ and\ \bibinfo {author} {\bibfnamefont {P.}~\bibnamefont
  {Sutcliffe}},\ }\href@noop {} {\emph {\bibinfo {title} {Topological
  solitons}}},\ Cambridge Monographs on Mathematical Physics\ (\bibinfo
  {publisher} {Cambridge University Press},\ \bibinfo {year}
  {2004})\BibitemShut {NoStop}%
\bibitem [{SLa()}]{SLaSi}%
  \BibitemOpen
  \href {http://slasi.rpd.univ.kiev.ua} {}\bibinfo {note} {\texttt{SLaSi}
  spin--lattice simulations package code available at
  \url{http://slasi.rpd.univ.kiev.ua}}\BibitemShut {NoStop}%
\bibitem [{Note2()}]{Note2}%
  \BibitemOpen
  \bibinfo {note} {We take the integration time step $\Delta t = 0.025 \omega
  _0^{-1}$ for the accuracy $\Delta S_\protect \text {max} = \protect \qopname
  \relax m{max}\limits _{i, n} |S_{i, n}^4-S_{i, n}^5| < 0.01$, where $\omega
  _0 = JS^2/\hbar $, $i=x,y,z$ and superscript for $S_{i, n}$ indicates
  integration order. We used modified RKF45 scheme, where the time step is
  changed only if accuracy goal for $\Delta S_\protect \text {max}$ is not
  reached for avoiding of noise from step changing.}\BibitemShut {Stop}%
\bibitem [{Note3()}]{Note3}%
  \BibitemOpen
  \bibinfo {note} {Supplementary video \protect \url
  {http://slasi.rpd.univ.kiev.ua/sm/pylypovskyi12}}\BibitemShut {NoStop}%
\bibitem [{\citenamefont {Khodenkov}(2010)}]{Khodenkov10}%
  \BibitemOpen
  \bibfield  {author} {\bibinfo {author} {\bibfnamefont {G.}~\bibnamefont
  {Khodenkov}},\ }\bibfield  {title} {\enquote {\bibinfo {title} {Exchange
  reduction of the magnetization modulus in the vicinity of a bloch point},}\
  }\href {http://dx.doi.org/10.1134/S1063784210050245} {\bibfield  {journal}
  {\bibinfo  {journal} {Technical Physics}\ }\textbf {\bibinfo {volume} {55}},\
  \bibinfo {pages} {738--740} (\bibinfo {year} {2010})},\ \bibinfo {note}
  {10.1134/S1063784210050245}\BibitemShut {NoStop}%
\bibitem [{\citenamefont {Reinhardt}(1973)}]{Reinhardt73}%
  \BibitemOpen
  \bibfield  {author} {\bibinfo {author} {\bibfnamefont {J.}~\bibnamefont
  {Reinhardt}},\ }\bibfield  {title} {\enquote {\bibinfo {title}
  {Gittertheoretische behandlung von mikromagnetischen singularitseten},}\
  }\href
  {http://books.google.com.ua/books?ei=smTVTpH0HIqJhQeZ45ln&ct=result&hl=uk&id=jj9XAAAAYAAJ&dq=Reinhardt+International+journal+of+Magnetism&q=Reinhardt#search_anchor}
  {\bibfield  {journal} {\bibinfo  {journal} {Int. J. Magn.}\ }\textbf
  {\bibinfo {volume} {5}},\ \bibinfo {pages} {263} (\bibinfo {year}
  {1973})}\BibitemShut {NoStop}%
\bibitem [{Note4()}]{Note4}%
  \BibitemOpen
  \bibinfo {note} {\protect \url {http://cluster.univ.kiev.ua}}\BibitemShut
  {NoStop}%
\bibitem [{Note5()}]{Note5}%
  \BibitemOpen
  \bibinfo {note} {\protect \url {http://icybcluster.org.ua/}}\BibitemShut
  {NoStop}%
\end{thebibliography}
%
%

%

\end{document}